\documentclass[aps,pre,twocolumn,showpacs,superscriptaddress,groupedaddress]{revtex4-1}
\pdfoutput=1
\usepackage{hyperref}
\usepackage{epsfig}
\usepackage{amsmath}
\usepackage{graphicx}
\usepackage{wrapfig}
\usepackage{dcolumn}
\usepackage{bm}
\usepackage{amssymb}
\usepackage{titlesec}
\usepackage{mathtools}
\usepackage{relsize}
\usepackage{cleveref}
\usepackage[usenames,dvipsnames]{color}
\usepackage{ulem}

\begin{document}
\title{Epidemic Extinction Paths in Complex Networks}
\author{Jason Hindes}
\author{Ira B. Schwartz}
\affiliation{U.S. Naval Research Laboratory, Code 6792, Plasma Physics Division, Nonlinear Systems Dynamics Section, Washington, DC 20375}
\begin{abstract}
We study the extinction of long-lived epidemics on finite complex networks induced by intrinsic noise. Applying analytical techniques to the stochastic Susceptible-Infected-Susceptible model, we predict the 
distribution of large fluctuations, the most probable, or optimal path through a network that leads to a disease-free state from an endemic state, and the average extinction time in general configurations. Our predictions agree with Monte-Carlo simulations on several networks, including synthetic weighted and degree-distributed networks with degree correlations, and an empirical high school contact network. In addition, our approach quantifies characteristic scaling patterns for the optimal path and distribution of large fluctuations, both near and away from the epidemic threshold, in networks with heterogeneous eigenvector centrality and degree distributions. 

\end{abstract}
\maketitle

\section{\label{sec:INTRO} INTRODUCTION}
Understanding the dynamics of infectious processes in complex networks is an important problem, both in terms of generalizing concepts in statistical mechanics and applying them to public health \cite{Pastor,Anderson,Dorogovtsev,Hindes1,Vespignani1}. A primary question in infectious disease modeling is how to control an outbreak, with the ultimate goal of reducing the number of individuals able to spread infection to zero. This process, by which an epidemic is extinguished, is called extinction or disease fade-out \cite{Doering,Schwartz, Meerson2,Meerson3,Assaf2}. To understand and possibly achieve extinction, mathematical models can be useful, where extinction can be naturally captured in terms of a dynamical transition from an endemic state (e.g., fluctuating equilibrium or cycle) to a disease-free state \cite{Assaf1}. 

Although it is known that random fluctuations are the cause of extinction in finite populations, the process of extinction does not happen in the deterministic systems analyzed in the vast majority of works on endemic dynamics in networks -- where contacts between infectious and susceptible individuals are typically assumed to be well above an epidemic threshold or bifurcation point \cite{Pastor,Meerson3,Nasell}. Consequently network-control prescriptions often reduce to bringing systems below a bifurcation point \cite{VaccRev,Schwartz,Drakopoulos}. One may ask, is targeting sub-threshold regimes as a control method necessary or even optimal? In actuality the spread of disease is a highly stochastic process both in terms of the natural randomness inherent in contact processes and fluctuations due to time-varying and uncertain environments \cite{Meerson3,Meerson}. These stochastic effects make extinction inevitable, even above threshold, in finite networks and should be reflected in epidemic controls \cite{Schwartz2,Billings}. In fact, recent work has shown that optimal control of networks with noisy dynamics leverages randomness and a network's natural, noise-induced pathway between distinct states \cite{Hindes1,Motter}. Continuing in this line of thinking, we seek a prescription for computing epidemic extinction pathways through complex networks.   

Such issues have received much attention in well-mixed and spatially homogeneous models \cite{Dykman,Meerson,Kamenev,Billings,Schwartz2,Meerson4}. It has been demonstrated in many works that noise and a system's dynamics can couple in such a way as to induce a large fluctuation -- effectively driving a system from one state to another \cite{Schwartz3}. If the fluctuation is a {\it rare event} in the weak noise limit, then the process is captured by a path that is a maximum in probability, or optimal path (OP), where all others are exponentially less likely to occur. The formalism borrows from analytical mechanics, describing the OP as a {\it least-action} trajectory in some effectively classical system, and allows one to predict the dynamical extinction pathway and the average time needed to realize it \cite{Dykman2, Friedlin}. Some recent works have made progress in understanding extinction in networks, (e.g., deriving bounds for average extinction times), but do not make use of the path-based formalism outlined here \cite{Durrett,Mountford,Van,Munoz,Buono,Assaf3}. 

The following layout of the paper describes epidemic extinction through complex networks with intrinsic (demographic) noise in terms of large fluctuations and rare-event theory. Sec.\ref{sec:WKB} constructs the formalism: combining  a mean-field approximation for endemic dynamics on networks with a Wentzel--Kramers--Brillouin $(WKB)$ technique that allows for an analytical description of the distribution of large fluctuations and the OP. The limiting form of the OP is discussed near the epidemic threshold in Sec.\ref{sec:Threshold}, and away from threshold in Sec.\ref{sec:Scaling}. Sec.\ref{sec:SOLU} addresses how to compute the OP, extinction time, and their dependencies in certain cases, including in networks with large spectral gaps (Sec.\ref{sec:SpecGap}) and degree distributions (Sec.\ref{sec:DegreeDist}). Throughout, predictions are compared to real and synthetic network simulations.  

\section{\label{sec:WKB} LARGE FLUCTUATIONS, MEAN-FIELD,  AND WKB APPROXIMATION}
In order to predict epidemic extinction in general contact networks it is necessary to consider an arbitrary weighted adjacency matrix, $A$, where 
each element, $A_{ij}$, represents the strength of a link, or contact, from node $i$ to node $j$, in a graph with $N$ nodes. Given this representation, a network's epidemic dynamics, assuming a simple Susceptible-Infectious-Susceptible Markov process (SIS) is captured by the states and transitions of its nodes; i.e., node $i$ is either ``infected", denoted $\nu_{i}\!=\!1$, or ``susceptible", $\nu_{i}\!=\!0$. Furthermore, node $i$ changes its state $\nu_{i}\!:\!0\rightarrow\!1$ with probability per unit time $\beta(1-\nu_{i})\sum_{j} A_{ij}\nu_{j}$, and $\nu_{i}\!:\!1\rightarrow\!0$ with probability per unit time $\alpha\nu_{i}$, where $\beta$ and $\alpha$ are known as the infection and recovery rates, respectively \cite{Pastor,Hindes1,Vespignani,Vespignani1}. Since the elements of $A$ are proportional to probabilities, $A$ is assumed to be nonnegative. It is important to note that there is inherent noise in the SIS model defined, which arises from the underlying {\it stochastic} reactions \cite{Gillespie}.

In order to analyze the stochastic dynamics, it is useful to consider an ensemble consisting of $C$ identical networks with the same $A$, but {\it independent} realizations of the stochastic dynamics \cite{Ott2}. Each node can be specified by a graph position $i$, and ensemble number $c$, with state $\nu_{i,c}\!\in\!\{0,1\}$. In this way, the number of infected nodes in the ensemble with graph position $i$ is $I_{i}\!=\!\sum_{c} \nu_{i,c}$, with corresponding transitions and rates: $I_{i}\!\rightarrow\!I_{i}+1$ with rate $R_{i}^{+}(\bold{I})\!\equiv\!\beta\sum_{c}(1-\nu_{i,c})\sum_{j} A_{ij}\nu_{j,c}\!=\!\beta\sum_{j}A_{ij}\sum_{c}(1-\nu_{i,c})\nu_{j,c}$, and $I_{i}\!\rightarrow\!I_{i}-1$ with rate $R_{i}^{-}(\bold{I})\!\equiv\!\alpha I_{i}$. To simplify our analysis, it is useful to make a {\it mean-field} approximation and replace $\nu_{i,c}$ by the ensemble average, $I_{i}/C$: $R_{i}^{+}(\bold{I})\!\approx\!\beta\sum_{j}A_{ij}I_{j}(1-I_{i}/C)$, so that the transition rates depend explicitly on $\bold{I}=\left<I_{1},I_{2},...,I_{N}\right>$ alone. This approximation neglects correlations between neighboring graph positions \cite{Goltsev,Mata,Pastor,FN5}. Ultimately, we are interested in the limit of large $C$, so that $x_{i}\!\equiv\!I_{i}/C$ gives a continuous fraction of infected nodes, or density, in graph position $i$. In this way, the large ensemble allows us to consider continuous densities even in discrete networks with unique graph positions.  

Given the stochastic reactions and rates $R_{i}^{+}$ and $R_{i}^{-}$,  the ensemble dynamics is described by a probability distribution, $P(\bold{I},t)$, satisfying a master equation:
\begin{align}
&\frac{\partial P}{\partial t}(\bold{I},t)=\sum_{i}R^{+}_{i}(\bold{I}-\bold{1}_i)P(\bold{I}-\bold{1}_i,t)-R^{+}_{i}(\bold{I})P(\bold{I},t) \nonumber \\
&+R^{-}_{i}(\bold{I}+\bold{1}_i)P(\bold{I}+\bold{1}_i,t)-R^{-}_{i}(\bold{I})P(\bold{I},t),
\label{eq:MasterEquation}
\end{align}
where $\bold{1}_{i}=\left<0{}\;_{1},0{}\;_{2},...,1{}\;_{i}, 0{}\;_{i+1},...\right>$. Because extinctions in large networks ($N\!\gg\!1$) with long-lived epidemics are rare events with small probabilities, we are interested in the tails of $P(\bold{I},t)$, where $\bold{I}$ corresponds to a large deviation from the average behavior, and is accompanied by an {\it exponential} reduction in probability. 

\begin{figure}[h]
\includegraphics[scale=0.33]{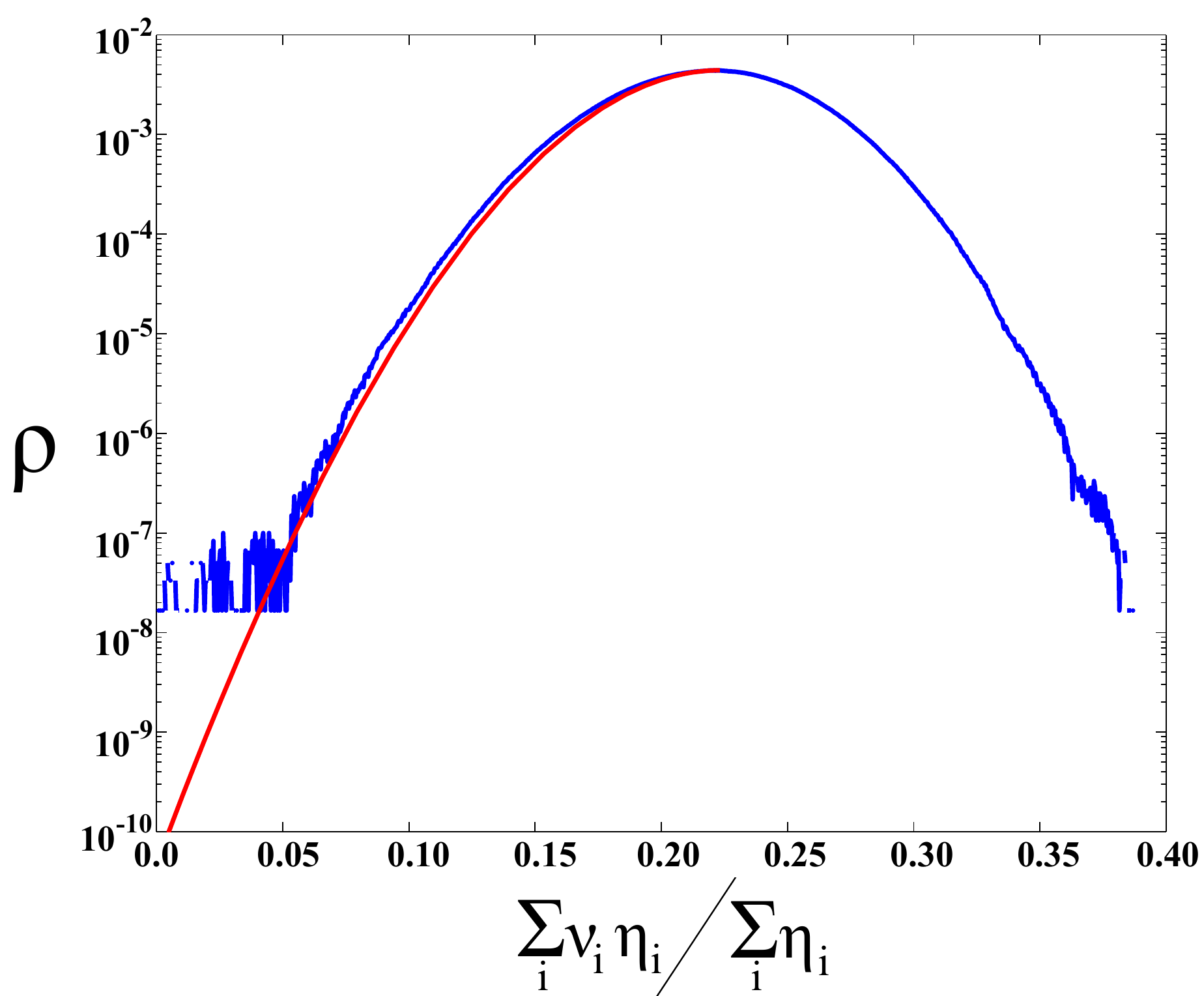}
\caption{(Color online) Histogram of a single stochastic realization of the SIS model on a high school contact network (HS). The probability (blue) is shown versus the average infection weighted by eigenvector centrality: $\eta_{i}$ for node $i$. Predictions for a single realization from Eq.(\ref{eq:Distribution}) are shown in red. Network details are given in Sec.\ref{sec:SOLU}.}
\label{fig:Distribution}
\end{figure}

This intuition suggests looking for solutions of Eq.(\ref{eq:MasterEquation}) with an exponential, or $WKB$ form, $P(\bold{I},t)=ae^{-CS(\bold{x},t)}$ \cite{Assaf1,Dykman2}. The WKB solution for the ensemble distribution can be viewed as a product of independent and identical distributions for each realization in the ensemble. Hence, we can approximate the probability distribution of states for a single realization, $\rho(\boldsymbol{\nu},t)$, by
\begin{align}
\label{eq:Distribution}
\rho(\boldsymbol{\nu},t)\cong\rho(\bold{x},t)=be^{-S(\bold{x},t)}. 
\end{align}
Predictions from Eq.(\ref{eq:Distribution}) (when combined with Eqs.(\ref{eq:EOM1}-\ref{eq:Action}) below) are in good agreement with simulations on an empirical high school network \cite{Salathe}, shown in red in Fig.\ref{fig:Distribution}. 

We can find the leading contribution to $P(\bold{I},t)$ by substituting the WKB ansatz into Eq.(\ref{eq:MasterEquation}), expanding in powers of the small parameter $1/C$ (e.g., $S(\bold{x}\pm\bold{1}_i/C)\!\approx\!S(\bold{x})\pm(1/C)\partial S/\partial x_{i}$), and neglecting terms of $\mathcal{O}(1/C)$ or smaller, where $C\!\gg\!1$. This approximation converts the master equation into a Hamilton-Jacobi equation (HJE):
\begin{align}
\label{eq:HJE}
\partial S/\partial t+H(\bold{x},\partial S/\partial\bold{x})=0,
\end{align}
where $S$ and $H$ are called the Action and Hamiltonian, respectively. The latter is a function of the infection density at graph position $i$, $x_{i}$, and its conjugate momentum, $p_{i}=\partial S/\partial x_{i}$:
\begin{align}
\!\!H(\bold{x},\bold{p})\!=\!\sum_{i}\!\!\Bigg[\!\beta\big(1\!-\!x_{i}\big)\!\big(e^{p_{i}}\!-\!1\big)\!\!\sum_{j}\!A_{ij}x_{j}+\alpha x_{i}\big(e^{-p_{i}}\!-\!1\big)\!\Bigg]\!. 
\label{eq:Hamiltonian}
\end{align}

Just as in analytical mechanics, a convenient approach for solving the HJE is to solve Hamilton's equations of motion, $\dot{x}_{i}=\partial H/\partial p_{i}$ and $\dot{p}_{i}=-\partial H/\partial x_{i}$:
\begin{align}
\label{eq:EOM1}
\dot{x}_{i}&=\!\tilde{\beta}(1-x_{i})e^{p_{i}}\!\sum_{j}\!A_{ij}x_{j}-x_{i}e^{-p_{i}},\\
\label{eq:EOM2}
\dot{p}_{i}&= \!\tilde{\beta}\sum_{j}\!A_{ij}x_{j}\big(e^{p_{i}}\!-\!1\big)\!-\!A_{ji}\big(1\!-\!x_{j}\big)\!\big(e^{p_{j}}\!-\!1\big)\!-\!e^{\!-p_{i}}\!+\!1,
\end{align}
expressed in terms of the ratio $\tilde{\beta}\!\equiv\!\beta/\alpha$, and the time, $\tau\!\equiv\!\alpha t$. Crucially, solutions of the HJE extremize $S$, when expressed as the integral
\begin{align}
S(\bold{x},t)=\int_{\bold{x}\small{(t=t_{0})}}^{\bold{x}}\!\!\bold{p}\cdot{d\bold{x}}-\int_{t_{0}}^{t}H(\bold{x},\bold{p})dt',
\label{eq:Action}
\end{align} 
where $\bold{x}(t)$ and $\bold{p}(t)$ are determined from Eqs.(\ref{eq:EOM1}-\ref{eq:EOM2})\cite{Dykman2}. Because $S$ is minimized, the probability of the corresponding trajectory is maximized-- a consequence of the WKB approximation. Therefore, all that is needed to find the most probable path to extinction (OP), and $\rho(\bold{x},t)$ (Eq.(\ref{eq:Distribution}) and Eq.(\ref{eq:Action})) is the appropriate solution of Eqs.(\ref{eq:EOM1}-\ref{eq:EOM2}). Such solutions can be determined from boundary conditions, and computed as detailed in Sec.\ref{sec:SOLU}.
 
\begin{figure*}[t]
\centering
\includegraphics[scale=0.256]{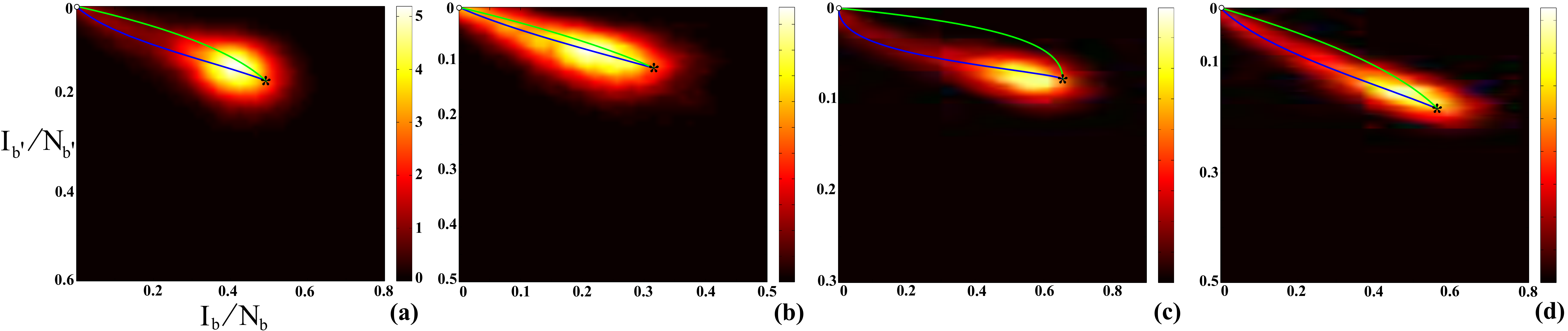}
\caption{(Color online) Pre-history density of the final $N$ events that ended in extinction in $400$ realizations of the SIS model on several fixed networks. The stochastic trajectories were projected into the fraction of infected nodes with high $(b)$ and low $(b')$ eigenvector centrality. Predicted paths are shown in blue for comparison (from the endemic state($*$) to extinction($\circ$)), and contrasted with the path into the endemic state, $p_{i}\!\equiv\!0$ (green). (a) WBA network: $\tilde{\beta}\lambda\!=\!1.88$, $\eta_{b}\!\geq\!0.050$, and $0.014\!\leq\!\eta_{b'}\!\leq\!0.018$ (see Sec.\ref{sec:SpecGap}). (b) HS network: $\tilde{\beta}\lambda\!=\!1.34$, $\eta_{b}\!\geq\!0.052$, and $0.0111\!\leq\!\eta_{b'}\!\leq\!0.021$. (c) Positively correlated bimodal network ($PC$): $\tilde{\beta}\lambda\!=\!2.8$, $w\!=\!0.23$, $N\!=\!500$, $k_{1}\!=\!5$, and $k_{2}\!=\!50$ (where $k_{1}$ and $k_{2}$ are degrees and $w$ is a degree-correlation parameter explained in Sec.\ref{sec:DegreeDist} and Sec.\ref{sec:AppB}). (d) Negatively correlated bimodal network ($NC$): $\tilde{\beta}\lambda\!=\!1.9$, $w\!=\!-0.26$, and $N\!=\!400$; high and low-centralities imply $k\!\geq\!40$ and $k\!\leq\!10$, respectively, for (c) and (d). Note: a position on the heat map with color ``$5$" means that five times as many of the total $400N$ events crossed the position as compared to a position with a color ``$1$".}
\label{fig:NetworkExtinctHeat}
\end{figure*}

From inspection of Fig.\ref{fig:Distribution} we notice that $\rho(\bold{x},t)$ is a maximum at the endemic equilibrium ($\bold{x}\!=\!\bold{x}^{*}$), implying $\partial S/\partial\bold{x}\!=\!\bold{0}$ and a boundary condition: $\dot{\bold{x}}\!=\!\bold{0}$, $\dot{\bold{p}}\!=\!\bold{0}$, $\bold{x}\!=\!\bold{x}^{*}$, and $\bold{p}\!=\!\bold{0}$. Second, at the extinct state $(\bold{x}\!=\!\bold{0})$ the distribution has {\it negative} slope, $\partial S/\partial\bold{x}\!<\!\bold{0}$. Furthermore, $\rho(\bold{x},t)$ is approximately time-independent, or {\it quasi-stationary} \cite{Dykman2,Assaf1,Assaf2}. In the $WKB$ ansatz, we have $\partial S/\partial t\!=\!H\!=\!0\; \forall t$. Therefore the final boundary condition is $\dot{\bold{x}}\!=\!\bold{0}$, $\dot{\bold{p}}\!=\!\bold{0}$, $\bold{x}\!=\!\bold{0}$, and $\bold{p}\!=\!\bold{p}^{*}$ \cite{Assaf1,Schwartz3,Dykman}. OPs computed with Eqs.(\ref{eq:EOM1}-\ref{eq:EOM2}) and the stated boundary conditions are compared with stochastic trajectories ending in extinction for several networks in Fig.(\ref{fig:NetworkExtinctHeat}).

Two important details should be pointed out. Since the distribution is time-independent, and therefore ``zero energy", $S(\bold{x})$ is simply the line integral of the momentum along the OP, from Eq.(\ref{eq:Action}). Also, the $\bold{p}\equiv\bold{0}$ solution of Eqs.(\ref{eq:EOM1}-\ref{eq:EOM2}) gives the familiar {\it quenched mean-field} equations for SIS model on complex networks. Therefore, the WKB approach generalizes mean-field results to include large fluctuations.     

\begin{figure}[h]
\includegraphics[scale=0.38]{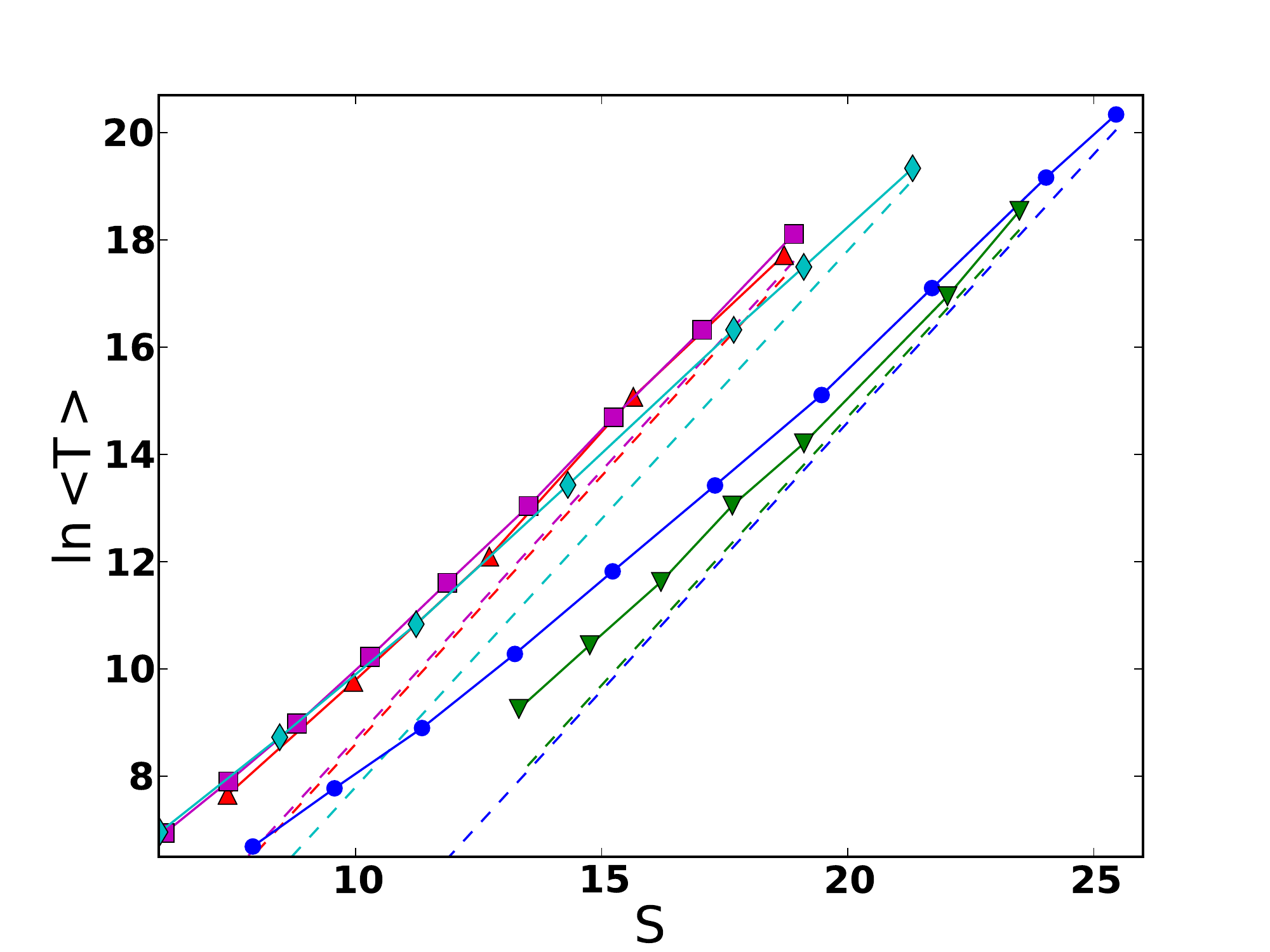}
\caption{(Color online) Log of the average extinction times vs. Actions, Eq.(\ref{eq:Action}), for several networks: WBA$(\circ)$; HS$(\Box)$; $PC$ $(\triangledown)$; $NC$ $(\triangle)$; CM network with $N\!=\!1000$ and a degree distribution, $g(k)\!=\!k^{-2.5}\!/\!\sum_{k'=20}^{500}k'^{-2.5}$ $(\diamondsuit)$. Average times are taken from at least $400$ stochastic realizations of the SIS dynamics on a single fixed network, and are shown with symbols. Dashed lines show the expected scaling $\ln{\!\left<T\right>}\sim S+\text{const.}$}
\label{fig:ScalingTimes}
\end{figure}

In general, by studying Eqs.(\ref{eq:EOM1}-\ref{eq:EOM2}) we can learn how a network's infection density is coupled to its large fluctuations -- together generating the most likely transition sequence through a network leading to extinction. In addition to the distribution of large fluctuations, Eq.(\ref{eq:Distribution}), an important observable from the above formalism is the geometry of the OP; e.g., specifying the shape of infection density in the different graph positions as a network makes its way from a large epidemic to extinction. Examples are shown in Fig.\ref{fig:NetworkExtinctHeat} and Fig.\ref{fig:Scaling}. Another important observable is the average extinction time for a given network, $\left<T\right>$, which is expected to take the form:
\begin{align}
 \left<T\right>=B(\tilde{\beta},A)e^{S(\bold{x}=\bold{0})}/\alpha,
\label{eq:AvgTime}
\end{align}
from the assumption that absorption into the extinct state has a rate, or inverse time, proportional to the probability \cite{Doering,Meerson,Assaf1,Schwartz3}. For sufficiently large $S$, the exponential contribution dominates, and therefore $\left<T\right>\!\sim\!e^{S(\bold{0})}$ (as demonstrated in Fig.(\ref{fig:ScalingTimes}) for several networks \cite{FN1}). We note that beyond a theoretical interest, the framework presented can be augmented with control strategies designed to minimize the Action, Eq.(\ref{eq:Action}), thus producing exponential and optimal reductions in the lifetime of epidemics on networks \cite{Hindes1}. 

\subsection{\label{sec:Threshold} Near threshold behavior}
Since the OPs are in a high $2N$-dimensional space, they must be found by numerically solving the two-point boundary value problem in general, given Eqs.(\ref{eq:EOM1}-\ref{eq:EOM2}). However, analytic properties can be derived in certain limiting cases, which are useful for guiding intuition and for initializing algorithms (see Sec.(\ref{sec:SpecGap})). An important case discussed in this section is for $\tilde{\beta}$ just above the epidemic threshold, or transcritical bifurcation, $\tilde{\beta}_{c}\!\equiv\!1/\lambda^{(1)}$ -- where $\lambda^{(1)}$ is the largest eigenvalue of $A$ \cite{Pastor,Goltsev,Mata}. At this point the endemic and extinct state meet, and below which no long-lived epidemic occurs. 

In order to describe the path when $\tilde{\beta}\!\gtrsim\!\tilde{\beta}_{c}$, it is useful to assume that $\tilde{\beta}\!=\!(1+\delta)/\lambda^{(1)}$, with $\delta \ll 1$, and first find the equilibria as functions of $\delta$. Substituting the series $x_{i}^{*}\!=\!\sum_{n}\delta^{n}x_{i,n}$ for each $i$ and $\bold{p}\!=\!\bold{0}$ into $d\bold{x}/d\tau\!=\!\bold{0}$, and collecting powers of $\delta$ (e.g., to $\mathcal{O}(\delta^{2})$) gives:
\begin{align}
\label{eq:NearThresholdX1}
\lambda^{(1)} x_{i,1}&=\sum_{j}A_{ij}x_{j,1},\\
\label{eq:NearThresholdX2}
\lambda^{(1)} x_{i,2}&=\sum_{j}A_{ij}[x_{j,2}+(1-x_{i,1})x_{j,1}].
\end{align} 
Furthermore, by decomposing Eqs.(\ref{eq:NearThresholdX1}-\ref{eq:NearThresholdX2}) into the eigenbasis of $A$, $x_{i,n}=\!\sum_{m=1}^{N}G_{n}^{(m)}\eta_{i}^{(m)}$, where $\boldsymbol{\eta}^{(m)}$ is the $m$th right eigenvector of $A$ with eigenvalue $\lambda^{(m)}$, and taking the inner product, $\sum_{i}\zeta_{i}x_{i}$, of Eqs.(\ref{eq:NearThresholdX1}-\ref{eq:NearThresholdX2}) with the left eigenvector, $\boldsymbol{\zeta}^{(1)}$, we find $\bold{x}^{*}$ to $\mathcal{O}(\delta)$. A similar procedures gives $\bold{p}^{*}$, when $\bold{x}\!=\!\bold{0}$:
\begin{align}
\label{eq:Delta1}
x_{i}^{*}&= \delta \eta_{i}^{(1)}/\sum_{j}\zeta^{(1)}_{j}{\eta_{j}^{(1)}}^{2}+\mathcal{O}(\delta^{2}),\\
\label{eq:Delta2}
p_{i}^{*}&= -\delta \zeta_{i}^{(1)}/\sum_{j}\eta^{(1)}_{j}{\zeta_{j}^{(1)}}^{2}+\mathcal{O}(\delta^{2})
\end{align}  
(assuming the normalization $\sum_{i}\zeta^{(1)}_{i}\eta^{(1)}_{i}\!=\!1$). Examining Eqs.(\ref{eq:Delta1}-\ref{eq:Delta2}), we see that $x_{i}^{*}$ and $p_{i}^{*}$ are proportional to the principal right and left eigenvectors of $A$ near the bifurcation, respectively. In particular, if $\boldsymbol{\eta}^{(1)}$ and $\boldsymbol{\zeta}^{(1)}$ contain relatively few nodes that are significantly large compared to most others, we expect the infection density and fluctuations to be localized around these nodes\cite{Goltsev}. 
 
Further insight on the effects of topology near threshold can be gained by considering the Action along the path, Eq.(\ref{eq:Action}). To this end, it is useful to introduce a length parameter, $a\in[0,1]$, so that we can express the coordinates as $x_{i}(a)\!\approx\!x_{i}^{*}(1-a)$ and $p_{i}(a)\!\approx\!p_{i}^{*}a$, where the linear form is the simplest satisfying the boundary conditions to $\mathcal{O}(\delta)$ (see Fig.\ref{fig:Scaling} insets). Integrating over the path gives the action near bifurcation, $S(a)\!=\!\sum_{i}\int_{0}^{a}{p_{i}(a')da'(dx_{i}/da')}$:
\begin{align}
S(\bold{x}(a))=\frac{\delta^{2}\left(a-a^{2}/2\right)}{\sum_{j}\zeta^{(1)}_{j}{\eta_{j}^{(1)}}^{2}\sum_{l}\eta^{(1)}_{l}{\zeta_{l}^{(1)}}^{2}}+\mathcal{O}(\delta^{3}). 
\label{eq:Action2}
\end{align}

We note that Eq.(\ref{eq:Action2}) is interesting, since the known expression for the complete graph is generalized by a topological factor that depends on the moments of the centrality distribution. Typically, as the distribution becomes broad, the topological factor in Eq.(\ref{eq:Action2}) is reduced, such that the Action differs significantly from the limiting case, $\eta_{i}\!=\!\zeta_{i}\!=\!1/\sqrt{N}\Rightarrow S=N\delta^{2}(a-a^{2}/2)$ \cite{Assaf1}. This is intuitive, since for heterogeneous networks, infection is most prevalent around a comparatively small number of nodes, who must recover without reinfection in order for extinction to occur. The effects of heterogeneous eigenvector centrality are explored in more detail in Sec.\ref{sec:SOLU}. 
\section{\label{sec:SOLU}SPECIAL SOLUTIONS}
In general, the OP is of interest away from threshold. However, since the OP is a heteroclinic connection of Eqs.(\ref{eq:EOM1}-\ref{eq:EOM2}), in practice it must be constructed numerically, e.g., through shooting, or quasi-newton methods, etc. For example, the paths shown in Fig.(\ref{fig:NetworkExtinctHeat}) were found from an iterative action minimizing method (IAMM) \cite{Lindley2}. In the IAMM, OPs are generated from a least-squares algorithm that minimizes the residuals between Eqs.(\ref{eq:EOM1}-\ref{eq:EOM2}) and finite-difference approximations. The boundary conditions specified in Sec.\ref{sec:WKB} are used to close the differencing. Often the small $\delta$ limit, Eqs.(\ref{eq:Delta1}-\ref{eq:Delta2}), can be used as an initial guess. However the dimension for the minimization is $2Nd$ where $d$ is the number of discrete points in the differencing and $N$ is the size of the network, which is prohibitively large for large $N$. Therefore, in practice it is necessary to coarse-grain the network in some way.  Two such approaches are discussed in the following sections for networks with large spectral gaps (Sec.\ref{sec:SpecGap}) and specified degree distributions (Sec.\ref{sec:DegreeDist}).  

\subsection{\label{sec:SpecGap} Large spectral gaps}
In general, $A$ can be usefully expanded in terms of its eigenvalues and eigenvectors: $A_{ij}\!=\!\sum_{n=1}^{N}\lambda^{(n)}\eta^{(n)}_{i}\zeta^{(n)}_{j}$. Of particular interest, for strongly connected graphs, $\boldsymbol{\eta}^{1}$ and $\boldsymbol{\zeta}^{1}$ are positive and unique, and $\lambda^{(1)}$ is equal to the spectral radius, by the Perron-Frobenius theorem. Moreover, in many strongly connected networks of interest, it is the case that $\lambda^{(1)}\!\gg\!\lambda^{(n)}$, with large spectral gaps, and therefore, $A_{ij}\!\approx\!\lambda^{(1)}\eta^{(1)}_{i}\zeta^{(1)}_{j}$. In such cases, $A$ can be coarse-grained along a single dimension as demonstrated below.    

Given a large spectral gap, a simple coarse-graining is to bin $\boldsymbol{\eta}^{(1)}$, assuming a number of bins, $B$, and a distribution, $f_{b}$, for the number of nodes in a given bin, $b$. In the following, we assume that $A$ is symmetric, so that $\boldsymbol{\eta}^{(n)}\!=\!\boldsymbol{\zeta}^{(n)}$. In this case, nodes can be ordered according to increasing $\eta_{l}^{(1)}$. The binning procedures follows: starting with the first node, the first bin is filled with nodes sequentially until the number of nodes equals $Nf_{1}$; then, the second bin is filled, etc. Once all nodes are binned, the dimension of Eqs.(\ref{eq:EOM1}-\ref{eq:EOM2}) can be reduced by replacing $\eta_{l}^{(1)}$, 
$x_{l}$, and $p_{l}$ with their bin averages $\forall l\!\in\!b$: $\eta_{b}\!\equiv\!\sum_{l\in b}\eta_{l}^{(1)}\!/(Nf_{b})$, $x_{b}\!\equiv\!\sum_{l\in b}x_{l}/(Nf_{b})$, $p_{b}\!\equiv\!\sum_{l\in b}p_{l}/(Nf_{b})$.  
This gives the following approximations to Eqs.(\ref{eq:EOM1}-\ref{eq:EOM2}) with reduced dimension $2B$:
\begin{align}
\label{eq:EOM_binned}
\dot{x}_{b}=&\;\tilde{\beta}\lambda^{(1)}\eta_{b}(1-x_{b})e^{p_{b}}\!\sum_{b'}\!Nf_{b'}\eta_{b'}x_{b'}-x_{b}e^{-p_{b}},\\
\dot{p}_{b}=&\;\tilde{\beta}\lambda^{(1)}\eta_{b}\!\sum_{b'}\!Nf_{b'}\eta_{b'}\big[x_{b'}\!\big(e^{p_{b}}\!-\!1\big)\!-\!\big(1\!-\!x_{b'}\big)\!\big(e^{p_{b'}}\!-\!1\big)\big]\!  \nonumber \\
&-\!e^{\!-p_{b}}\!+\!1.\!
\end{align} 
A final requirement is needed to ensure that the binned and original system have the same bifurcation point. We choose to renormalize $\eta_{b}$ so that $\sum_{b}\eta_{b}^{2}f_{b}N\!=\!\sum_{i}\eta_{i}^{2}\!=\!1$.

\begin{figure}[t]
\includegraphics[scale=0.23]{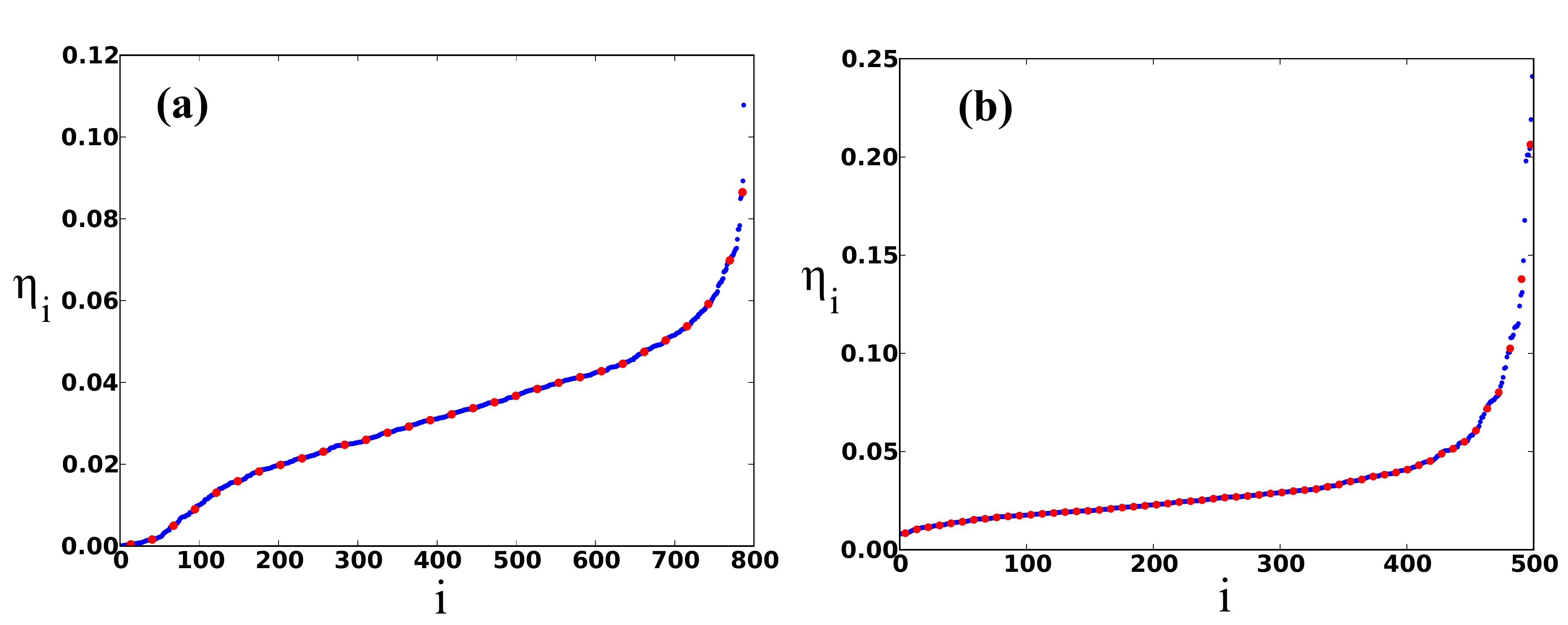}
\caption{(Color online) Example binnings of (a) HS and (b) WBA networks (Sec.\ref{sec:SpecGap}). Eigenvector centralities (blue) are shown for all nodes and compared with the average centrality in each bin (red). (a) 30 bins (b) 55 bins.}
\label{fig:Binning}
\end{figure}

The above procedure was applied to three networks (considered in Figs.\ref{fig:NetworkExtinctHeat}-\ref{fig:ScalingTimes}), where the bin distribution was assumed to be uniform for simplicity, $f_{b}\!=\!1/B$;  $B$ was chosen large enough so that the binned centralities closely matched the original (as in Fig.\ref{fig:Binning}), but not so large to preclude using $200\!-\!1000$ discretization points along the $OP$ in the IAMM. The first network in Fig.\ref{fig:NetworkExtinctHeat}(a) is a weighted Barab\'{a}si-Albert graph (WBA) with $N\!=\!500$ and initial degree for each node, $m\!=\!7$ \cite{Albert}. Every link was given a random weight, independently drawn from a uniform distribution over the range $[0,10]$ after the network was generated from the standard Barab\'{a}si-Albert algorithm ($\lambda^{(1)}\!=\!122, \lambda^{(2)}\!=\!58.6$). The second network in Fig.\ref{fig:NetworkExtinctHeat}(b) is a high-resolution American high school contact network (HS) with $788$ individuals and links representing close proximity interactions during the course of a day (measured using wireless motes \cite{Salathe}). Weights associated with links correspond to contact durations ($\lambda^{(1)}\!=\!6715, \lambda^{(2)}\!=\!4882$). Binning results for the WBA and HS networks are shown in Fig.\ref{fig:Binning} with $B\!=\!55$ and $B\!=\!30$, respectively. The third network was generated from a configuration model, $CM$, with $N\!=\!1000$ and a degree distribution, $g(k)\!=\!k^{-2.5}\!/\!\sum_{k'=20}^{500}k'^{-2.5}$, where $k$ is the number of links for a given node ($\lambda^{(1)}\!=\!80.9, \lambda^{(2)}\!=\!16.3$)\cite{Newman2}; $B\!=\!55$ for the CM network's OP computation. A related method for computing OPs through networks with specified degree distributions is discussed in Sec.\ref{sec:DegreeDist}, and was applied to the networks in Fig.\ref{fig:NetworkExtinctHeat}(c)-(d).
\subsubsection{\label{sec:Scaling}  Scaling with broad centrality distribution}
\begin{figure}[t]
\includegraphics[scale=0.23]{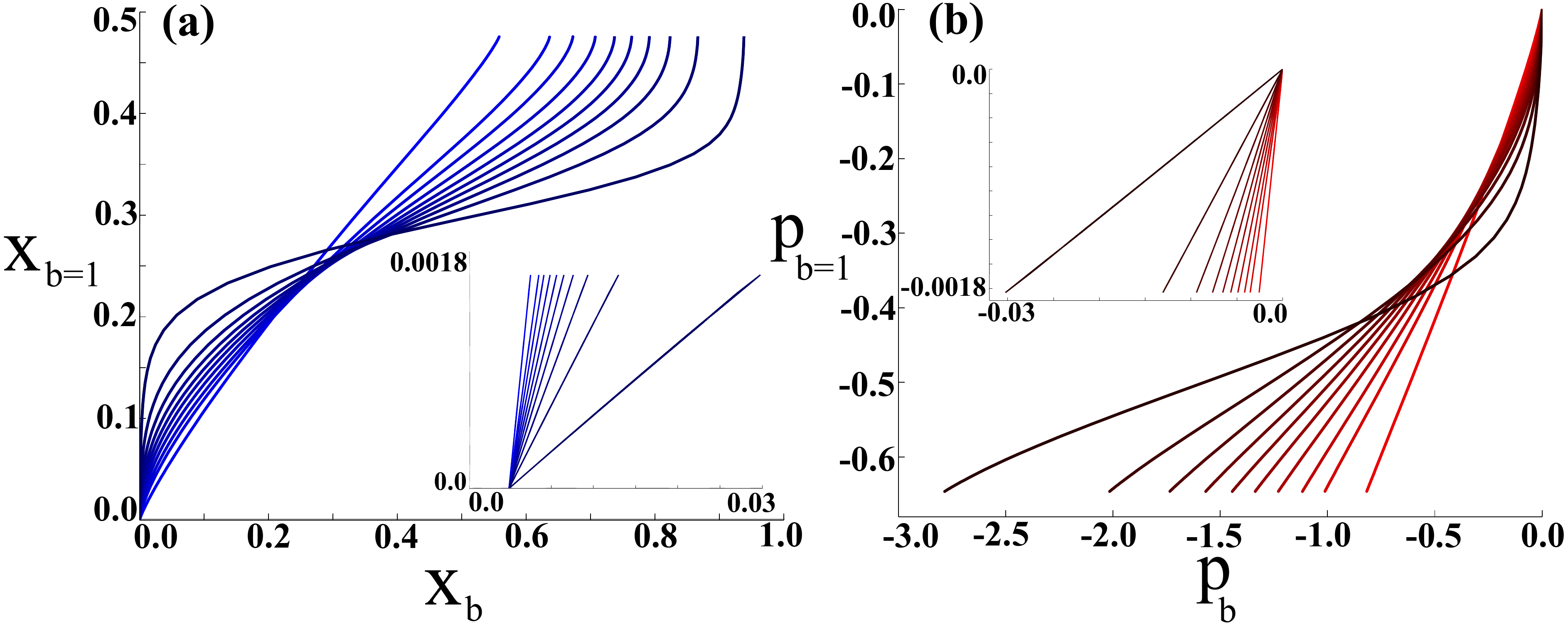}
\caption{(Color online) OP away from threshold ($\tilde{\beta}\lambda\!=\!7$) for WBA network projected into the (a) densities of infected nodes and (b) conjugate momenta for the lowest centrality bin vs.$\;$higher centrality bins (Sec.\ref{sec:SpecGap}). Path projections become increasingly dark as the bin number increases: $b=10,15,...55$ (increasing eigenvector centrality), where $B\!=\!55$. Insets show OPs for comparison when $\tilde{\beta}\lambda\!=\!1.02$.}
\label{fig:Scaling}
\end{figure}

As $\tilde{\beta}$ is increased above $\tilde{\beta}_{c}$, infection increases across the network. In particular, infection density can become high even at low centrality nodes. In this case the path to extinction through a network is more complex as its global dynamical structure becomes apparent. Example optimal paths for a WBA network when $\tilde{\beta}\!\gg\!\tilde{\beta}_{c}$ are shown in Fig.\ref{fig:Scaling}. We can see that a multi-step structure is visible in the relative change of infection density at different graph positions, which can be compared with the insets ($\tilde{\beta}\!\gtrsim\!1/\lambda^{(1)}$) \cite{Hindes1}. Though the path has a more complicated form away from threshold, some characteristic scalings can be captured in this region of parameter space for networks with heterogeneous eigenvector centralities (e.g., $f_{b}\sim\eta_{b}^{-\gamma}$) and where the large spectral gap approximation holds. Our approach in this section is to study the unstable and stable linear modes of $(\bold{x},\bold{p})$ near the endemic and extinct states, respectively, for such networks. These modes approximate the OP (a heteroclinic connection) near the equilibria, and are useful for describing how large fluctuations depend on centrality\cite{FN6}.   

\begin{figure}[t]
\raggedleft{\includegraphics[scale=0.23]{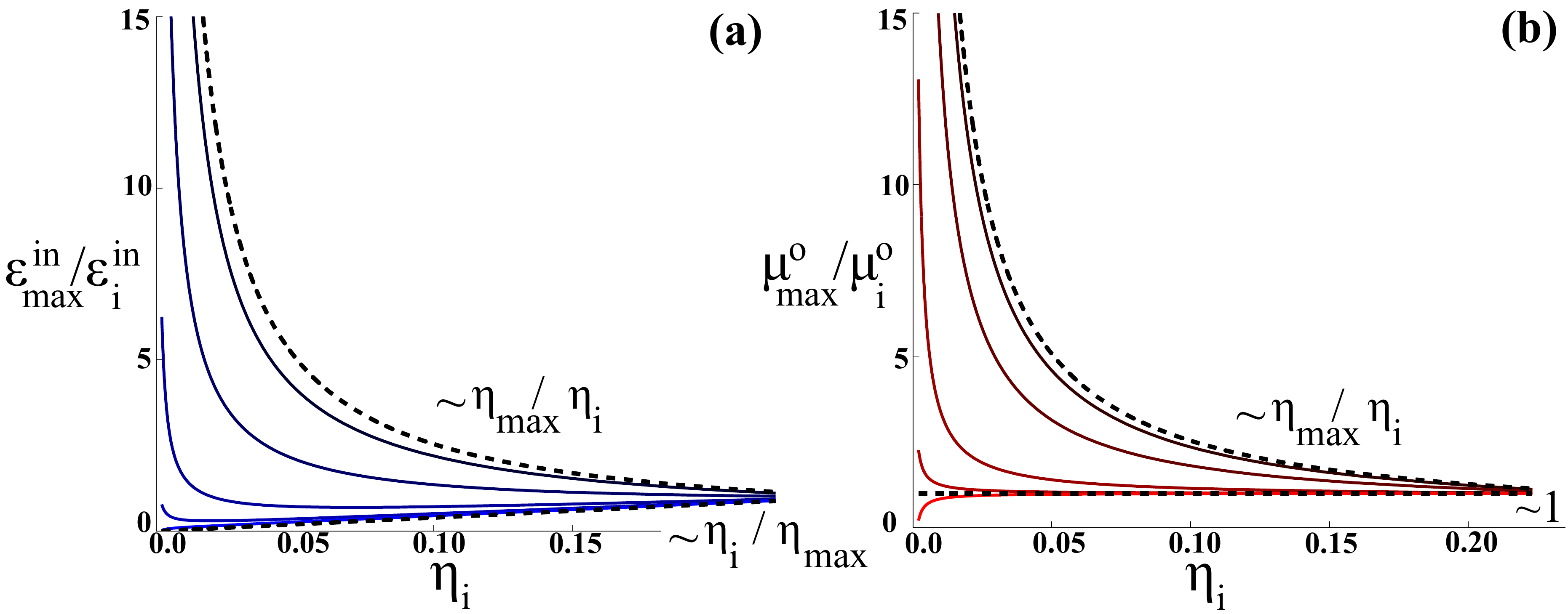}}
\caption{(Color online) Scaling of dynamical eigenmodes for nodes in a $WBA$ network relative to the maximum centrality value (max). (a) Solutions of Eqs.(\ref{eq:SigmIn}-\ref{eq:EpsIn}). (b) Solutions of Eqs.(\ref{eq:SigmaO}-\ref{eq:MuO}). Upper/lower dashed lines represent the predicted scaling near/away from threshold. Solid lines become increasingly light in color as $\tilde{\beta}\lambda$ increases: $\tilde{\beta}\lambda\!=\!1.05,1.25,2.0,2.85, \text{and}\; 3.35$.}
\label{fig:Modes}
\end{figure}

With this end in mind, we consider the dynamics of $(x_{i},p_{i})=(x_{i}^{*}+\epsilon_{i}^{o},\mu_{i}^{o})$ and $(x_{i},p_{i})=(\epsilon_{i}^{in},p_{i}^{*}+\mu_{i}^{in})$, for small $\boldsymbol{\epsilon}$ and $\boldsymbol{\mu}$, given $A_{ij}\approx\lambda^{(1)}\eta^{(1)}_{i}\eta^{(1)}_{j}$ (below, we drop the superscript (1) in $\boldsymbol{\eta}$ and $\lambda$ for convenience). Similar to Sec.\ref{sec:Threshold}, when $\tilde{\beta}\!\gtrsim\!\tilde{\beta}_{c}$, it is straightforward to show that $\epsilon_{i}^{o}, \epsilon_{i}^{in}, \mu_{i}^{o}$ and $\mu_{i}^{in}$ are simply proportional to $\eta_{i}$. Fig.\ref{fig:Modes} shows centrality scalings for the principal linear eigen-modes of Eqs.(\ref{eq:EOM1}-\ref{eq:EOM2}) near the equilibria. The upper dashed lines demonstrate the predicted scaling $\epsilon_{i}^{in}/\epsilon_{j}^{in}\!\sim\!\eta_{i}/\eta_{j}\!\sim\!\mu_{i}^{o}/\mu_{j}^{o}$ for a WBA network where the dark blue/red curves correspond to $\tilde{\beta}$ increasingly close to threshold ($\epsilon_{i}^{o}$ and $\mu_{i}^{in}$ scale similarly in this region). However, as $\tilde{\beta}$ is increased, shown in light blue/red, we can see that the scaling changes significantly \cite{Hindes1}.

In order to understand the change in scaling as $\tilde{\beta}$ is increased, we first consider the equilibria $x_{i}^{*}$ and $p_{i}^{*}$. Given the large spectral gap assumption, we find a simple form for each that is dependent on two parameters, $X$ and $P$:
\begin{align}
\label{eq:GapX}
x_{i}^{*}&= X\tilde{\beta}\lambda\eta_{i}/\big(1+X\tilde{\beta}\lambda\eta_{i}\big),\\
\label{eq:GapY}
p_{i}^{*}&=-\ln\!\bigg[1+\tilde{\beta}\lambda\eta_{i}\bigg(\!N\!\left<\eta\right>-P\!\bigg)\bigg] , 
\end{align}   
\noindent satisfying: $X\!=\!\sum_{j}\eta_{j}x^{*}_{j}$ and $P\!=\!\sum_{j}\eta_{j}e^{p_{j}^{*}}$, where $\left<\eta\right>$ is the average eigenvector centrality \cite{Ott2}. In particular, by assuming that infection densities are high in the endemic state at most  graph positions, i.e., $\tilde{\beta}\lambda\eta_{i}N\!\left<\eta\right>\!\gg\!1$, then $X\!\approx\!N\!\!\left<\eta\right>\!-\!1/[\tilde{\beta}\lambda\!\left<\eta\right>]$ and $P\!\approx\!1/[\tilde{\beta}\lambda\!\left<\eta\right>].$ Substituting these approximations into the linearized Eqs.(\ref{eq:EOM1}-\ref{eq:EOM2}) allows us to determine the dependence of the eigen-modes, $(\epsilon_{i}^{o}(t),\mu_{i}^{o}(t))\!=e^{t\sigma^{o}}\!(\epsilon_{i}^{o},\mu_{i}^{o})$ and $(\epsilon_{i}^{in}(t),\mu_{i}^{in}(t))\!=e^{t\sigma^{in}}\!(\epsilon_{i}^{in},\mu_{i}^{in})$, on $\eta_{i}$ near the equilibria.

Since infection densities are high near the endemic state away from threshold, we expect the most well connected nodes to be quickly reinfected after recovery, as compared to nodes that are less well connected. Therefore, we expect the OP out of the endemic state to correspond with an initial decrease in infection at low centrality positions. The scaling for this initial step is determined by the eigensolution of the linearized Eq.(\ref{eq:EOM2}) at $(x_{i}^{*},0)$: 
\begin{align}
\label{eq:SigmaO}
&1=\sum_{j}\tilde{\beta}\lambda\eta_{j}^{2}\big(1-x^{*}_{j}\big)\big/\big[1+\tilde{\beta}\lambda\eta_{j}X-\sigma^{o}\big],\\
\label{eq:MuO}
&\mu_{i}^{o}=\tilde{\beta}\lambda\eta_{i}\!\sum_{j}\eta_{j}\mu_{j}^{o}(1-x_{j}^{*})/\big[1+\tilde{\beta}\lambda\eta_{i}X-\sigma^{o}\big].
\end{align}    
In particular, $\sigma^{o}$ is positive and grows from zero with $\tilde{\beta}\!>\!\tilde{\beta}_{c}$ \cite{FN2}. When $\tilde{\beta}\lambda\eta_{i}N\!\left<\eta\right>\!\gg\!1$ and $f_{b}\!\sim\!\eta_{b}^{-\gamma}$ for large $\eta$, the summation in Eq.(\ref{eq:SigmaO}) $\sim\!\int\!\eta^{-\gamma+1}d\eta/(\eta-\text{const.})$, and converges in the limit of large maximum centrality, $\eta_{max}$, when $\gamma\!>\!2$. This implies that $\sigma^{o}$ does not depend sensitively on $\eta_{max}$ in this region and therefore we can consider the limit of large centrality in Eq.(\ref{eq:MuO}). By inspecting $\mu_{i}^{o}$ for large $\eta_{i},$ we see that it tends to a constant, i.e., $\mu_{i}^{o}/\mu_{j}^{o}\sim1$, (since the sum over $j$ is $i$-independent) in good agreement with numerical solutions of Eqs.(\ref{eq:SigmaO}-\ref{eq:MuO}) away from threshold-- shown in Fig.\ref{fig:Modes}(b)(light red). Interestingly, $\mu^{o}_{i}$ becomes largest for small $\eta_{i}$ (light red) and increases quickly to a constant for large $\eta_{i}$. Since the momenta are nearly equal across nodes in the network, the Action's derivatives w.r.t. infection density are nearly equal across nodes. A similar procedure gives the scaling $\epsilon_{i}^{o}/\epsilon_{j}^{o}\sim\eta_{j}/\eta_{i}$, which is found by expanding the linearized Eq.(\ref{eq:EOM1}) in $1/\tilde{\beta}\lambda\eta_{i}N\!\left<\eta\right>$, e.g., $x_{i}^{*}\!\approx\!1-1/\tilde{\beta}\lambda\eta_{i}N\!\left<\eta\right>$ (Sec.\ref{sec:AppA}). Therefore, the infection density at a given node decreases {\it inversely} proportional to its eigenvector centrality, i.e, the reciprocal scaling of the OP near threshold.

Following the same approach, the scaling of the OP near the extinct state is determined by the eigen-solution of the linearized Eq.(\ref{eq:EOM1}) at $(0,p_{i}^{*})$:
\begin{align}
\label{eq:SigmIn}
&1=\sum_{j}\tilde{\beta}\lambda\eta_{j}^{2}e^{p_{j}^{*}}\big/\big[\sigma^{in}+e^{-p_{j}^{*}}\big],\\
\label{eq:EpsIn}
&\epsilon_{i}^{in}=\tilde{\beta}\lambda\eta_{i}e^{p_{i}^{*}}\!\sum_{j}\eta_{j}\epsilon_{j}^{in}/\big[\sigma^{in}+e^{-p_{i}^{*}}\big].
\end{align}
In particular, $\sigma^{in}$ is negative, decreases from zero with $\tilde{\beta}\!>\!\tilde{\beta}_{c}$, and is similarly insensitive to large $\eta_{max}$. Hence, taking the limit of large $\eta_{i}$, given $e^{-p_{i}^{*}}\!\approx\!1\!+\!\beta\lambda\eta_{i} N\!\left<\eta\right>\![1\!-\!1\big/\beta\lambda\eta_{i} N\!\left<\eta\right>]$, implies $\epsilon_{i}^{in}/\epsilon_{j}^{in}\sim\eta_{j}/\eta_{i}$ in Eq.(\ref{eq:EpsIn})-- as found near the endemic state and shown in Fig.\ref{fig:Modes}(b)(light blue). However, in contrast to the behavior near $x_{i}^{*}$,  we find that $\mu_{i}^{in}$ increases with $\eta_{i}$, for small $\eta_{i}$, before reaching a constant, $\mu_{i}^{in}/\mu_{j}^{in}\sim1$ (see details in Sec.\ref{sec:AppA} and Fig.\ref{fig:Modes2}). The scalings near the extinct state imply that the last segment of the OP is coincident with a final recovery of residual infections at low-centrality nodes, while the momentum is largest at high centralities. 

Putting the scalings near the equilibria together, we can infer that infection density decreases rapidly in high-centrality graph positions at a boundary layer between the endemic and extinct states -- since we have shown that their change is small compared to low centralities near the equilibria  \cite{Hindes1,FN3}. This can be seen in Fig.\ref{fig:Scaling}(a), where projections of the OP into high and low-centralities, on the $x$ and $y$ axes respectively, show a characteristic pattern in which segments with large horizontal slope occur between two segments with large vertical slope. 

\subsection{\label{sec:DegreeDist} Degree distributions}
In addition to understanding the OP for a given network defined by $A$, it is useful to understand the qualitative structure of paths and Actions for networks with similar statistical properties \cite{Pastor,Hindes1,Dorogovtsev}. A popular approach is to consider networks with a specified distribution for the fraction of nodes with $k$ links, $g(k)$ (where $k$ is called the degree), which is the focus of this section. Often, additional information is stipulated, such as a degree-correlation function -- typically in the form of a specified probability that a link starting from a node with degree $k$ leads to a node with degree $k'$, $o(k' | k)$ \cite{Vespignani,Pastor2,Pastor3}. OPs for networks with such properties can be found by {\it approximating} $A$ given these distributions, and substituting into Eqs.(\ref{eq:EOM1}-\ref{eq:EOM2}).

As is customary, we replace $A_{ij}$ by its expectation value in the ensemble of simple networks with $g(k)$ and $o(k'|k)$, which is called the {\it annealed} network approximation. In particular, $A_{ij}$ is approximated by the probability that nodes $i$ and $j$ are connected, $A_{ij}\!\approx\!o(k_{j}|k_{i})k_{i}\big/Ng(k_{j})$, or the probability that node $i$ is connected to any node with degree $k_{j}$ along a single link, multiplied by the number of possible links, and divided by the number of nodes with degree $k_{j}$ \cite{Dorogovtsev,Pastor,Hindes1}. Note that for link consistency, $A_{ij}=A_{ji}$, the distributions must satisfy the constraint: $ko(k'|k)p(k)=k'o(k|k')p(k')$ \cite{Pastor3}. With this substitution for $A_{ij}$ into Eqs.(\ref{eq:EOM1}-\ref{eq:EOM2}), Hamilton's equations depend on the density of infection for nodes with the same degree $k$, $x_{k}$, and their momentum, $p_{k}$:
\begin{align}
\label{eq:EOMk1}
\dot{x}_{k}=&\tilde{\beta}k(1-x_{k})e^{p_{k}}\!\sum_{k'}\!o(k'|k)x_{k'}-x_{k}e^{-p_{k}},\\
\label{eq:EOMk2}
\dot{p}_{k}=&\tilde{\beta}k\sum_{j}\!o(k'|k)\!\left[x_{k'}\big(e^{p_{k}}\!-\!1\big)\!-\!\big(1\!-\!x_{k'}\big)\!\big(e^{p_{k'}}\!-\!1\big)\right]\! \nonumber \\
&-\!e^{\!-p_{k}}\!+\!1.
\end{align} 

Notably, Eqs.(\ref{eq:EOMk1}-\ref{eq:EOMk2}) reduce to the heterogeneous mean-field dynamics for networks when $p_{k}\!\equiv\!0$;  $p_{k}\!\neq\!0$ entails extinction in degree-correlated topologies with dimension of $(\bold{x},\bold{p})$ equal to twice the number of degree classes. The analysis and results for degree-distributed networks are analogous to Sec.\ref{sec:Threshold}-Sec.\ref{sec:Scaling}. For example, for degree-distributed networks the familiar proportionality of the Action on the number of nodes in $A$, is found from Eq.(\ref{eq:Action}) \cite{Dykman2}: 
\begin{align}
\label{eq:ActionK}
S(\bold{x})\!=\!N\sum_{k}g(k)\!\int_{x_{k}^{*}}^{x_{k}}\!{p_{k}dx'_{k}}.    
\end{align}
Moreover, with the appropriate substitution of the largest eigenvalue, $\lambda$, and the corresponding right eigenvector, $v_{k}$, of $ko(k'|k)$ in Eq.(\ref{eq:Action2}) \cite{Pastor3}, we find the Action at the extinct state for degree-correlated networks near the epidemic threshold:
\begin{equation}
\label{eq:TFdc}
S=\frac{1}{2}N\delta^{2}\frac{\left<v^{2}\right>^{3}}{\left<v^{3}\right>^{2}}+\mathcal{O}(\delta^{3}), 
 \end{equation}
where $\left<v^{n}\right>\!=\!\sum_{k}g(k)v_{k}^{n}$. 

Extinction paths and times for two example networks are shown in Figs.\ref{fig:NetworkExtinctHeat}-\ref{fig:ScalingTimes} \cite{Hindes1,Hindes4}. The networks have a bimodal degree-distribution with positive $(PC)$ and negative ($NC$) degree-correlations (Figs.\ref{fig:NetworkExtinctHeat}(c) and (d) respectively), where positive implies an increased probability relative to an uncorrelated network for nodes with similar degree to share an edge. Correlated bimodal networks can be constructed in a straightforward manner as detailed in Sec.\ref{sec:AppB}. Fig.\ref{fig:NetworkExtinctHeat} (c)-(d) shows OPs for example parameters computed from Eqs.(\ref{eq:EOMk1}-\ref{eq:EOMk2}), and projected into the densities of infected low and high-degree nodes. Qualitatively, we can see that OP projections into infection densities are significantly closer to lines with unit slope (which is the case for uncorrelated networks with small variance in $k$) in the $NC$ case (d), than for the $PC$, (c). 

The change in the OP's shape with correlations suggests a reduction/enhancement of the effects of network heterogeneity with negative/positive correlations. For positive correlation, infection is more prevalent around high-degree nodes. This is reflected in the principal eigenvectors of $ko(k'|k)$ for the two examples, where the low-degree component is $5.4$ times greater in the $NC$ (parameters in Fig.\ref{fig:NetworkExtinctHeat}). In fact the topological factor, $\left<v^{2}\right>^{3}\!\!/\!\left<v^{3}\right>^{2}$, is $2.2$ times greater for the $NC$, given the same $N$ and distance to bifurcation, $\delta\!=\!\tilde{\beta}\lambda\!-\!1\!\gtrsim\!0$. Therefore we expect the probabilities for large fluctuations to be smaller by the same power and extinction times to be larger by the same power. Equivalently, if comparing fixed extinction time, the $PC$ must be taken to larger $\tilde{\beta}\lambda$ and/or $N$. This is demonstrated in Fig.\ref{fig:ScalingTimes}, where the largest times shown correspond to $\tilde{\beta}\lambda\!=\!1.9$ and $N\!=\!400$ for the $NC$, and $\tilde{\beta}\lambda\!=\!2.8$ and $N\!=\!500$ for the $PC$.
\begin{figure}[t]
\includegraphics[scale=0.295]{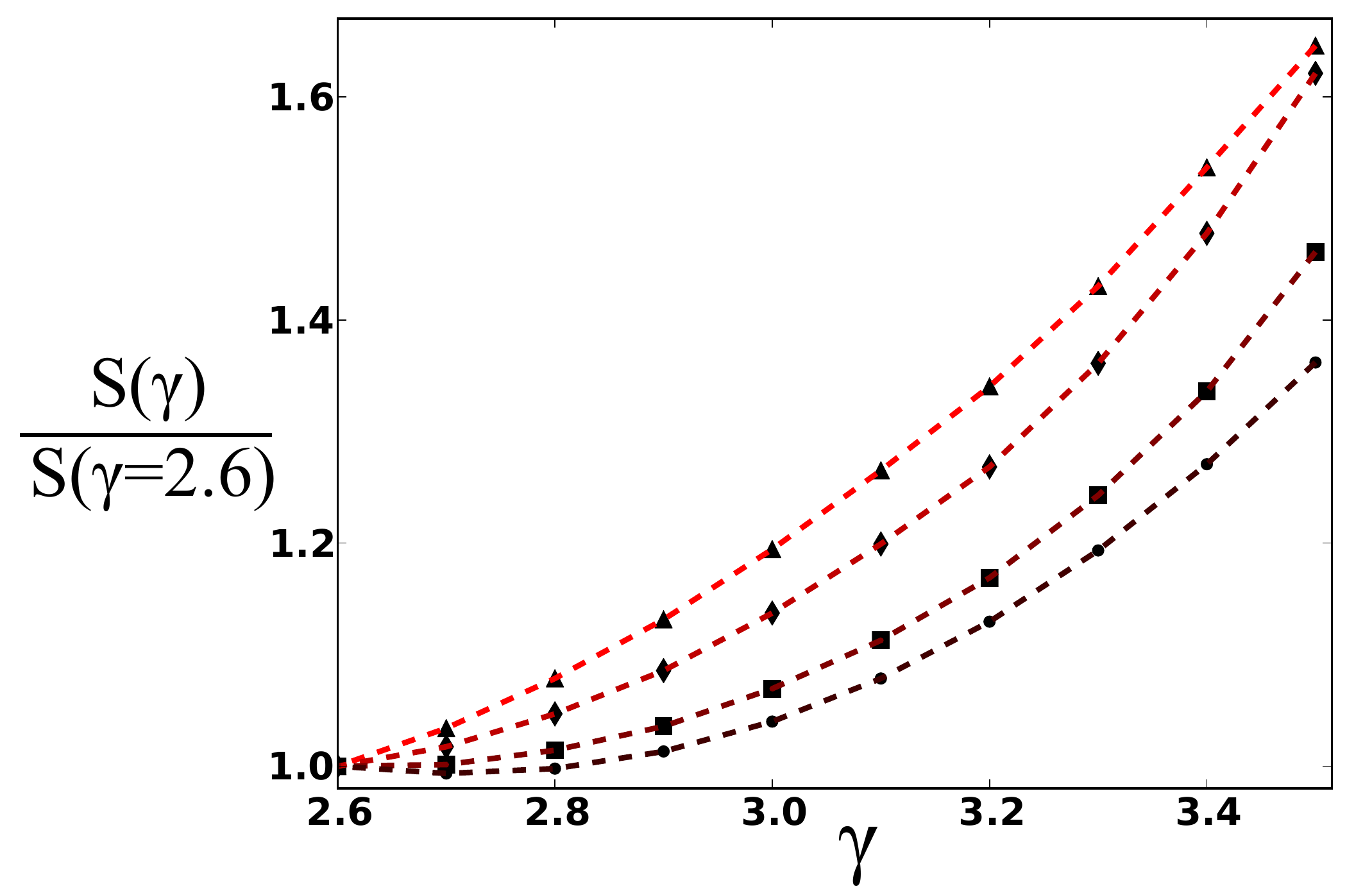}
\caption{(Color online) Relative Actions at the extinct state vs.$\;$degree-distribution exponent for truncated power-law networks: $g(k,\gamma)\!=\!k^{-\gamma}\!/\!\sum_{k'=20}^{400}k'^{-\gamma}$. $(\circ)$ $\tilde{\beta}\lambda\!=\!1\!+\!\delta, \delta\!\ll\!1$; $(\Box)$ $\tilde{\beta}\lambda\!=\!1.1$; $(\diamondsuit)$ $\tilde{\beta}\lambda\!=\!1.5$; $(\triangle)$ $\tilde{\beta}\lambda\!=\!2.0$. A bin width of $0.015$ was used to coarse-grain $kg(k)/\!\left<k\right>$ (Sec.\ref{sec:DegreeDist}).}
\label{fig:Action}
\end{figure} 

The above example raises an interesting question of how fluctuations and extinction times vary with statistical properties in a network, such as degree-heterogeneity, which can be anticipated from the network Action. A more realistic class of heterogenous networks have power-law degree distributions, $g(k)\!\sim\! k^{-\gamma}$, where the level of degree-heterogeneity grows with decreasing $\gamma$. Fig.\ref{fig:Action} shows the predicted Actions at the extinct state as a function of $\gamma$ for truncated and uncorrelated, $o(k|k')\!=\!kg(k)/\!\left<k\right>$, power-law distributions with several fixed distances to threshold. Interestingly, for such networks we can see that Actions vary as much as $60\%$ when ($\tilde{\beta}\lambda\!=\!2$), with broader distributions resulting in significantly smaller Actions, and therefore exponentially larger probabilities of large fluctuations and exponentially smaller extinction times. 

The Action curves  were found by solving Eqs.(\ref{eq:EOMk1}-\ref{eq:EOMk2}) with the boundary conditions specified in Sec.\ref{sec:WKB}, and computing Eq.(\ref{eq:ActionK}). In addition to computation, the lower black-curve in Fig.\ref{fig:Action} gives the analytic scaling near threshold, found from Eq.(\ref{eq:TFdc}), by substituting the eigenvector for uncorrelated random networks, $v_{k}\!=\!k\!\Rightarrow S\!=\!N\delta^{2}\!\left<k^{2}\right>^{3}\!\!/\!\left<k^{3}\right>^{2}$ \cite{Pastor3,Hindes1}. For the computed curves, it was useful to reduce the dimension for the IAMM by binning the distribution $o(k|k')\!=\!kg(k)/\!\left<k\right>$ with a similar procedure as Sec.\ref{sec:SpecGap}. Our approach, was to select a small bin width for $kg(k)/\!\left<k\right>$ (e.g., 0.015), and sequentially add degree classes to a bin, starting with the smallest $k$ and first bin, until the sum of $kg(k)/\!\left<k\right>$ over $k$ in a given bin equaled or exceeded the bin width. Then, the next bin was filled with the same bin width, etc. In the final step, degrees in Eqs.(\ref{eq:EOMk1}-\ref{eq:EOMk2}) were replaced by their bin's average and $o(k'|k)$ by the sum over $kg(k)/\!\left<k\right>$ in each bin \cite{Hindes1}.
\begin{figure}[t]
\includegraphics[scale=0.31]{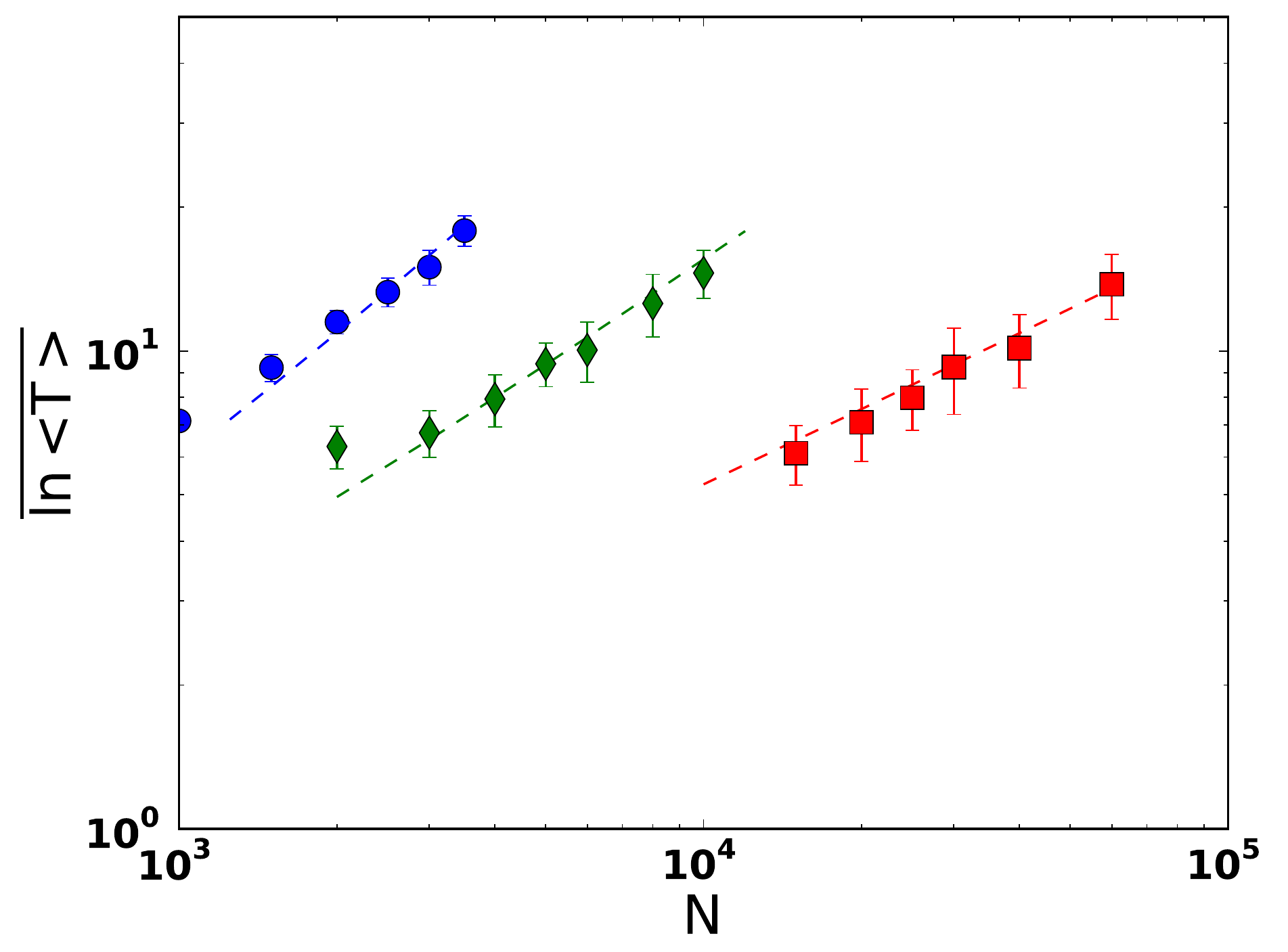}
\caption{(Color online) System-size scaling of the average log of the average extinction times for $CM$ networks with degree distribution $g(k,\gamma)\!=\!k^{-\gamma}\!/\!\sum_{k'=30}^{\infty}k'^{-\gamma}$: $\gamma\!=\!4.5$ $(\circ)$, $\gamma\!=\!3.7$ $(\diamondsuit)$, and $\gamma\!=\!3.4$, $(\Box)$, with $\tilde{\beta}\!\left<k^2\right>\!/\!\left<k\right>\!=\!1.18,$ $1.20,$ and $1.25,$ respectively \cite{FN5}. Dashed lines show the predicted scalings from Eq.(\ref{eq:Action3}). Average times were computed from at least 50 simulations for each network, and $ln\left<T\right>$ was averaged over 20 different networks. Networks were selected so that $k_{max}$ was within 10\% of $k_{min}N^{1/[\gamma-1]}$. Error bars represent the standard deviation of $ln\left<T\right>$.}
\label{fig:Ndep}
\end{figure} 

\subsubsection{\label{sec:Ndep} System-size scaling for modest N}
Another interesting feature of extinction times in power-law networks concerns their scaling with system-size when there is no truncation in $g(k)$. It is known for degree-homogeneous networks, such as simple-complete or Erd\H{o}s-R\'{e}nyi graphs, that the Action scales linearly with the system size \cite{Assaf1, Lindley}. Below we show that near threshold for {\it modest} $N$, a range of scalings are possible depending on the exponent, $\gamma$. As indicated above, the Action near threshold depends on a topological factor that is a function of the moments of $g(k)$, which can depend on $N$.

Here we continue to use the annealed network approximation, though for very large $N$ this is known to break down for random networks with unbounded degree as localization effects become important \cite{Goltsev}. Therefore, we restrict ourselves to $N$ and minimum degree, $k_{min}$, such that $\lambda\approx\left<k^{2}\right>\!/\!\left<k\right>$. When $\gamma\!>\!4$, $\left<k^{3}\right>$ is finite for power-law networks, i.e., independent of $N$ for large $N$, and thus $S\!\sim\!N$ near threshold, $\tilde{\beta}\!\gtrsim\!\left<k\right>\!/\!\left<k^{2}\right>$ \cite{Vespignani,Pastor2} -- though higher-order terms may be $N$-dependent for $\delta\!\ll\!1$, we expect them to grow more slowly than $N$ \cite{FN5}. On the other hand, if $\gamma\!<\!4$, $\left<k^{3}\right>$ is a function of the maximum degree, $k_{max}$, which follows a simple scaling: $k_{max}\sim k_{min}N^{1/[\gamma-1]}$, for a finite network with minimum degree $k_{min}$ \cite{Cohen}. The customary approach is to approximate the statistical moments of $g(k)$ given $k_{max}$, allowing one to find the scaling of $S$ with $N$. For example, when $k_{min}\!\gg\!1$, the discrete sum, $\left<k^{3}\right>\!=\!\sum_{k}g(k)k^{3}\approx\tilde{C}\int_{k_{min}}^{k_{max}}k^{3-\gamma}dk$. Introducing $\gamma\!=\!3\!+\!\alpha$ with $\alpha\!\in\!(0,1)$, we get $\left<k^{3}\right>\!\approx\! \tilde{C}k_{min}^{1-\alpha}N^{[1-\alpha]/[2+\alpha]}/[1-\alpha]$, where $\tilde{C}\!=\!k_{min}^{2+\alpha}[2+\alpha]$ (from normalization of $g(k)$). Computing the moments of $g(k)$ in this way, gives the Action at the extinct state to $\mathcal{O}(\delta^{2})$ from Eq.(\ref{eq:ActionK}):
\begin{align}
S=\frac{\delta^{2}}{2}\frac{(1-\alpha)^{2}(2+\alpha)}{\alpha^{3}}N^{1-2[1-\alpha]/[2+\alpha]},
\label{eq:Action3}
\end{align}

The above suggests that in the heterogeneous mean-field approximation, the $\mathcal{O}(\delta^{2})$ contribution to the Action can increase sub-linearly in $N$ for $\gamma\!\in\!(3,4)$ near threshold \cite{Assaf3} as suggested in Fig.\ref{fig:Ndep}. However for very large networks, and no truncation in $k_{max}$, eventually $\lambda\!\sim\!\text{max}\{\sqrt{k_{max}},\left<k^{2}\right>\!/\!\left<k\right>\}\!\gg\!1$, and the analysis presented is no longer valid, including the expansion in $\delta$. Moreover, there is some evidence for multiple epidemic thresholds in networks with unbounded $k_{max}$ as $N\rightarrow\infty$ \cite{Mata2}.
Since such issues are not yet resolved, we leave the description of extinction in very large networks with unbounded degree distributions, and the crossover between localized and delocalized extinction for future study.  


\section{\label{sec:CONC}CONCLUSION}
This work dealt with the extinction of long-lived endemic states above epidemic thresholds on static finite networks with infection dynamics given by the stochastic SIS model. The optimal path to extinction (OP), the distribution of large fluctuations, and the average extinction time were computed by combining mean-field and WKB-approximation techniques. The path-based formalism presented enabled us to predict extinction in general networked populations, and extract several of its intriguing signatures in complex topologies, including the multistep scaling of the OP in networks with heterogeneous eigenvector centrality, as well as an increase in the probability of large fluctuations with increased topological heterogeneity. Although theoretical in nature, the generality of our approach allowed us to consider several applications, including weighted empirical and degree-correlated topologies.  

Though the results show good qualitative and quantitative agreement with Monte-Carlo simulations in both real and synthetic networks, improved accuracy can be achieved in a straightforward manner by following our synthesized prescription, namely: Using as an ansatz in a network's master equation the exponential function of an {\it Action} (typically requiring some accurate mean-field approximation), and taking a large system-size limit. The result is a Hamilton-Jacobi equation that generates a dynamical system with twice the dimension of the mean-field. The OP can be found by solving the two-point boundary value problem of Hamilton's equations of motion beginning at an endemic state and ending at an extinct state, which define OP endpoints. Thus the theory changes the stochastic analysis of large fluctuations in networks to one that may be analyzed using a deterministic formalism whose zero-fluctuation limit is a mean-field theory. Furthermore, our approach can be more generally applied to other questions concerning noise and network dynamics, such as epidemic extinction in adaptive networks, switching in social networks, network inference in the presence of large fluctuations, and optimal control of networks with fluctuating dynamics \cite{Hindes1,Motter}.

 \section{\label{sec:Ack}ACKNOWLEDGMENTS}
\noindent J. H. is a National Research Council postdoctoral fellows. I.B.S was supported by the U.S. Naval Research Laboratory funding (N0001414WX00023) and office of Naval Research (N0001416WX00657) and (N0001416WX01643). We are very grateful to C. R. Myers and M. Assaf for useful discussions. 
\section{\label{sec:AppA}APPENDIX A}
\begin{figure}[t]
\includegraphics[scale=0.245]{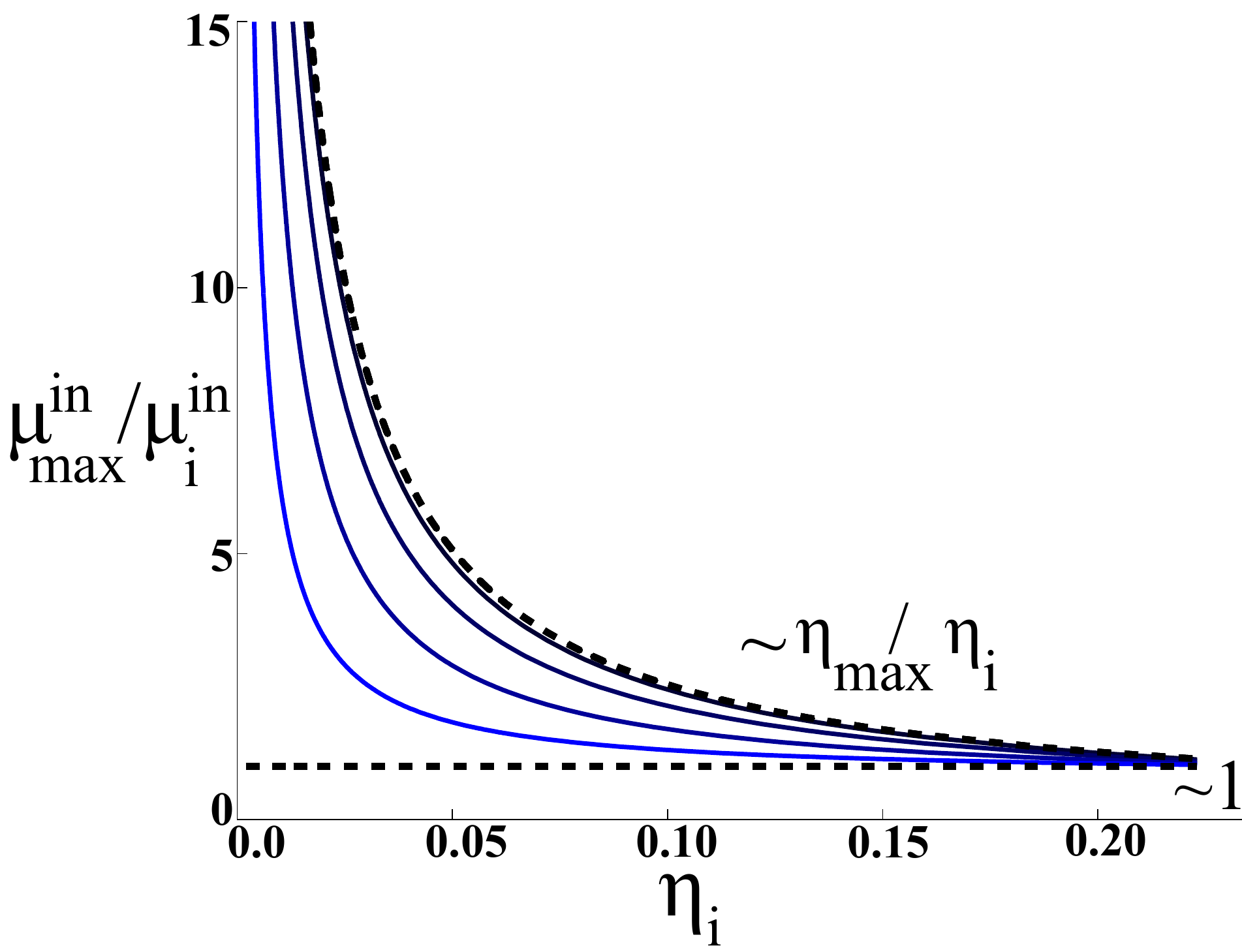}
\caption{Scaling of momentum near the extinct state for graph positions in a $WBA$ network relative to the maximum centrality value (max), corresponding to the dynamical mode Eq.(\ref{eq:EpsIn}). Upper/lower dashed lines represent the predicted scaling near/away from threshold. Solid lines become increasingly light in color as $\tilde{\beta}\lambda$ increases: $\tilde{\beta}\lambda\!=\!1.05,1.25,2.0,2.85, \text{and}\; 3.35$.\\}
\label{fig:Modes2}
\end{figure}

As described in Sec.\ref{sec:Scaling}, $\epsilon_{i}^{o}$ and $\mu_{i}^{in}$ satisfy the linearized Eqs.(\ref{eq:EOM1}-\ref{eq:EOM2}). When $\tilde{\beta}\lambda\eta_{i}N\!\left<\eta\right>\!\gg\!1$, the approximate linear systems are:  
\begin{align}
&\sum_{j}\frac{\eta_{j}\epsilon_{j}^{o}}{N\!\left<\eta\right>}\approx\;\epsilon_{i}^{o}\Big[\sigma^{o}\!+\!1\!+\tilde{\beta}\lambda\eta_{i}N\!\left<\eta\right>\!\Big(1\!-\!\big(1\big/\tilde{\beta}\lambda N\!\left<\eta\right>^{\!2}\!\big)\!\Big)\!\Big]\nonumber \\
&-\mu_{i}^{o}\Big[2\!-\!\big(1\big/\tilde{\beta}\lambda N\!\left<\eta\right>^{\!2}\!\big)\!-\!\big(1\big/\tilde{\beta}\lambda\eta_{i} N\!\left<\eta\right>\!\big)\Big].
\label{eq:EpsO}
\end{align}
\begin{align}
&\left[-1+\big((\sigma^{in}-1)\!\big/\tilde{\beta}\lambda\eta_{i} N\!\left<\eta\right>\!\big)\!\right]\!\mu_{i}^{in}\approx\;\!\!-\!\sum_{j}\!\frac{\eta_{j}\mu_{j}^{in}}{N\!\left<\eta\right>}\frac{1}{\tilde{\beta}\lambda\eta_{j}N\!\left<\eta\right>} \nonumber \\
&+\!\sum_{j}\frac{\eta_{j}\epsilon_{j}^{in}}{N\!\left<\eta\right>}\!\left[-2+(1\big/\tilde{\beta}\lambda\eta_{i}N\!\left<\eta\right>)+(1\big/\tilde{\beta}\lambda\eta_{j}N\!\left<\eta\right>)\right].
\label{eq:MuI}
\end{align}
Since the sums in Eqs.(\ref{eq:EpsO}-\ref{eq:MuI}) are independent of $i$, the limit of large $\tilde{\beta}\lambda\eta_{i}N\!\left<\eta\right>$, gives: $\epsilon_{i}^{o}/\epsilon_{j}^{o}\sim\eta_{j}/\eta_{i}$ and $\mu_{i}^{in}/\mu_{j}^{in}\sim1$. The latter can be seen in Fig.(\ref{fig:Modes2}).
However, as $\tilde{\beta}\!\gg\!\tilde{\beta}_{c}$ the continuous spectra,  $\sigma^{o}_{i}$ and $\sigma^{in}_{i}$,  for large $N$ of the linearized Eqs.(\ref{eq:EOM1}-\ref{eq:EOM2}) become relevant: $\epsilon_{i}^{in},\mu_{i}^{o}\sim\delta{(\eta-\eta_{i})}$, and 
\begin{align}
\sigma^{in}_{i}&=\tilde{\beta}\lambda{\eta_{i}}^{2}e^{p_{i}^{*}}-e^{-p_{i}^{*}},\\
\sigma^{o}_{i}&=1+\tilde{\beta}\lambda\eta_{i}\big[\sum_{j}\eta_{j}x_{j}^{*} -\eta_{i}(1-x_{i})\big]. 
\end{align}
This occurs as the denominators of Eq.(\ref{eq:SigmaO}) and Eq.(\ref{eq:SigmIn}) approach zero, and the single-mode analysis of Sec.\ref{sec:Scaling} is invalid. As a consequence, for very large $\tilde{\beta}$, the relevant modes directing the OP to extinction near the equillibria are extremely localized around low-centrality nodes.
\section{\label{sec:AppB}APPENDIX B}
Correlated bimodal networks can be constructed as follows. We assume that a fraction, $p$, of the network has high-degree near $k_{2}$ while the remaining nodes have low-degree near $k_{1}$. To build such networks, high-degree nodes are connected to each other with probability $k_{2}^{2}/[N\!\left<k\right>]\!+w$, where $w$ measures the assortativity above the uncorrelated construction and $\left<k\right>\!=\!(k_{1}(1-p)\!+\!k_{2}p)$. On the other hand, high and low-degree nodes are connected with probability $k_{1}k_{2}/[N\!\left<k\right>]-w$, and low-degree nodes are connected with probability $k_{1}^{2}/[N\!\left<k\right>]\!+\!w'$ -- where $w'$ is determined from the link-consistency constraint. In this way the degree distribution has two peaks centered around $k_{1}$ and $k_{2}$ as $N\rightarrow\infty$, and Eqs.(\ref{eq:EOMk1}-\ref{eq:EOMk2}) can be used to capture the OP and average extinction times assuming two degree classes with: $o(k_{2}|k_{2})\!=\!w+k_{2}p/\!\left<k\right>$, $o(k_{1}|k_{2})\!=\! -w+k_{1}(1-p)/\!\left<k\right>$, $o(k_{1}|k_{1})\!=\!-w'+k_{1}(1-p)/\!\left<k\right>$, and $o(k_{2}|k_{1})\!=\!-w'\!+\!k_{2}p/\!\left<k\right>$.


\begin{thebibliography}{9}
\bibitem{FN5} Techniques that include correlations can be used to significantly improve accuracy, but at the cost of higher dimensionality \cite{Mata}.
\bibitem{FN1} The deviation from the expected scaling is likely due to the pre-factor dependence $(B)$, the binning approximation (Sec.\ref{sec:SOLU}), and the inaccuracy of the mean-field assumption, particularly in the estimate for $\tilde{\beta_{c}}$ \cite{FN5}.
\bibitem{FN6} The usual stable and unstable modes of the endemic and disease-free states, respectively, (corresponding to $\bold{p}\!\equiv\!{0}$) are not discussed in Sec.\ref{sec:Scaling}.
\bibitem{FN3} Given the monotonic dynamics of the path, an upper bound for the scaling near the boundary layer is $dx_{i}/dx_{j}\sim \eta_{i}/\eta_{j}$ \cite{Hindes1}.
\bibitem{FN4} Due to the relatively small network sizes in Figs.\ref{fig:NetworkExtinctHeat}-\ref{fig:ScalingTimes}, the mean degrees for high and low-degree nodes in the bimodal networks (Sec.\ref{sec:AppB}) differ from the assumed values of $50$ and $5$. This introduces a quantitative error, which can be reduced by replacing $k_{1}$ and $k_{2}$ in  predictions by the measured mean degrees for the two node types.
\end{thebibliography}
\end{document}